\documentclass[aps,onecolumn,preprint,superscriptaddress,nofootinbib,floats]{revtex4}
\usepackage{amsmath,amssymb,color,mathrsfs, graphicx,verbatim,epsfig, bbm, wasysym}
\usepackage[hyperfootnotes=true]{hyperref}
\usepackage{slashed}
\allowdisplaybreaks

\def\lsim{\mathrel{\rlap{\lower4pt\hbox{\hskip1pt$\sim$}}
    \raise1pt\hbox{$<$}}}
\def\gsim{\mathrel{\rlap{\lower4pt\hbox{\hskip1pt$\sim$}}
    \raise1pt\hbox{$>$}}}

\newcommand{\be}{\begin{eqnarray}}
\newcommand{\ee}{\end{eqnarray}}
\newcommand{\njet}{N_{jets}}
\newcommand{\ttbar}{t\overline{t}}
\newcommand{\ttg}{\ensuremath{\ttbar\gamma}}

\def\addresses#1#2{\hbox to \hsize{\@tablebox{#1}\hfil\@tablebox{#2}}}
\def\@tablebox#1{\vtop{\hsize=5in \begin{flushleft} #1 \end{flushleft}}}

\def\beq{\begin{equation}}
\def\eeq{\end{equation}}
\def\bit{\begin{itemize}}
\def\eit{\end{itemize}}
\def\beqa{\begin{eqnarray}}
\def\eeqa{\end{eqnarray}}

\def\met{$\displaystyle{\not}E_T$}

\def\mtt{$m_{t\bar t}$}

\newcommand{\mo}{\mathcal{O}}

\newcommand{\invfb}{\rm fb^{-1}}

\begin{document}

\baselineskip 0.6cm

\begin{titlepage}

\thispagestyle{empty}

\begin{flushright}
\end{flushright}

\begin{center}

\vskip 2cm

{\Large \bf  Determining Top Quark Couplings at the LHC: Snowmass White Paper  }

\vskip 1.0cm
{\large  Jahred Adelman$^{1}$, Matthew Baumgart$^{2}$, Aran Garcia-Bellido$^{3}$, Andrey Loginov$^{1}$}
\vskip 0.4cm
{\it $^1$ Department of Physics, Yale University, New Haven, CT 06511} \\
{\it $^2$ Department of Physics, Carnegie Mellon University, Pittsburgh, PA 15213} \\
{\it $^3$ Department of Physics and Astronomy, University of Rochester, Rochester, NY 14627}\\
\vskip 1.2cm

\end{center}

\noindent Top quarks are a prime system for hunting for new physics.  Nonetheless, two decades on from their discovery few of their couplings
have been measured to high precision.  We present an overview of current determinations and the expected sensitivities with 300 and 3000~fb$^{-1}$ 
of 14 TeV LHC data.  In addition to direct limits on the top quark's renormalizable couplings to Standard Model bosons, we also explore what bounds 
can be set on the coefficients of higher-dimension operators, taking particular four-fermion operators that do not interfere with QCD as a test case.  
Every coupling we consider will benefit greatly from a dedicated study at the future LHC.  Some measurements, like the irrelevant operators, are systematics-limited 
and will saturate in the near-term.  Others, like the important $t \bar t H$ coupling, involve rare processes and thus demand as much data as possible.

\end{titlepage}

\setcounter{page}{1}

\section{Introduction}
\label{sec:intro}

The top quark has long been watched as a harbinger of new physics.  Its large coupling to the electroweak symmetry breaking sector motivates the precision study of its interactions with the Higgs, as well as the SU(2)$\times$U(1) gauge bosons.  As a further consequence of its large mass, the top quark's short lifetime ($\tau = 3.3\times 10^{-25}$ s) makes its weak decay distinctly observable.  Despite the challenges inherent in top reconstruction, new physics affecting the general quark sector may be more apparent in top quarks than in channels with purely light quark jets.  

To account for new physics, we can parametrize the couplings of top quarks in a couple different ways.  One method is the {\it vertex-function} or {\it form-factor} approach.  We write the interaction of the top quarks with Standard Model bosons and allow for anomalous terms.  For example, the coupling of a single top quark to the $W$~boson takes the form
\beq
\mathcal{L}_{int} \supset - \frac{g}{\sqrt{2}} \bar{b}\gamma^\mu \left( c^W_L P_L + c^W_R P_R\right)\,t\,W^-_\mu 
- \frac{g}{\sqrt{2}} \,\bar{b}\, \frac{i\sigma^{\mu\nu} q_\nu}{M_W} \left( d^W_L P_L + d^W_R P_R\right) \, t\, W^-_\mu \,+\, h.c., 
\label{eqn:anom}
\eeq
where $q$ is the $W$ momentum, $c^W_L \approx 1$, and $c^W_R,\, d^W_L,\, d^W_R$ vanish at tree level in the Standard Model (SM).  In general these $c,d$ couplings are functions of $q^2$.  This method has the advantage of transparency.  Additionally, in scenarios with new physics at low scales, we may have to know the full functional dependence  of the form factors with respect to $q^2$, and the {\it effective field theory} (EFT) method, based on an expansion in $q^2$, breaks down.  However, when there is a parametric separation between the scale of new physics and top quarks, EFT is the preferable approach.  

Instead of modifying renormalizable couplings, EFT changes the SM lagrangian by adding irrelevant operators, $\mo_i$, giving an effective lagrangian,
\beq
{\cal L}^{\rm eff} = \mathcal{L}_{SM} + \sum \frac{c_i}{\Lambda^2}\,\mathcal{O}_i \,+\, \cdots 
\eeq
The $c_i$ are the operators' dimensionless coefficients and $\Lambda$ is the scale of new physics.  The operators we add respect the gauge and global symmetries of the SM and are theoretically consistent for use in loop calculations.\footnote{One can find a more detailed description of the advantages of the EFT approach in \cite{Zhang:2010px}.}  Furthermore, they allow for interactions that cannot be written as vertex functions, such as a four-fermion operators, which we will study further.  Lastly, if an expansion in powers of $q^2/\Lambda^2$ is possible, then the vertex function approach can be subsumed by effective field theory.

This work presents a collection of analyses designed to probe the sensitivity of future experimental data sets to measuring top quark couplings.  In Section \ref{sec:fourfermi}, we place bounds on the coefficients of four-quark operators (two light quarks, two top quarks).  This demonstrates the scales of new physics we can probe in the top quark sector with high-luminosity LHC14 datasets. Section~\ref{vtb} describes the projected sensitivity of the LHC to the $Wtb$ coupling and its interpretation for searches for anomalous couplings or new particles.  Section \ref{sec:ttzero} examines the determination of the top quark's couplings to $\gamma$ and $Z$.  Section~\ref{sec:yukawa} discusses some prospects for top quark-Higgs boson physics at the LHC.

\section{Non-interfering Four-Quark Operators}
\label{sec:fourfermi}

There are many parametrizations of the irrelevant interactions that affect the top quark sector in the literature ({\it e.g.~}Refs.~\cite{AguilarSaavedra:2008zc,Zhang:2010dr,Adelman:2013gis}) with $\mo(10)$s of possible dimension-six operators.  A systematic study of the limits on the entire basis would be of great interest, but is beyond the scope of this work.  Constraints on many higher dimension operators can be found in Refs.~\cite{Greiner:2011tt,Drobnak:2011aa,Zhang:2012cd}.  We will focus on eight parity-odd operators, a set that does not interfere with QCD.  In addition to being heretofore unconstrained, one can search for them via a simple counting experiment.  They are
\begin{align}
\mo^{1,8}_{AV} &= (\bar q \, \gamma^\mu \gamma^5 \, T^{1,8} q ) (\bar t \, \gamma_\mu \, T^{1,8} t ), & \mo^{1,8}_{VA} &= (\bar q \, \gamma^\mu \, T^{1,8} q ) (\bar t \, \gamma_\mu \gamma^5 \, T^{1,8} t ), \nonumber \\
\mo^{1,8}_{PS} &= (\bar q \, \gamma^5 \, T^{1,8} q ) (\bar t \, T^{1,8} t ), & \mo^{1,8}_{SP} &= (\bar q \, T^{1,8} q ) (\bar t \, \gamma^5 \, T^{1,8} t ), 
\label{eq:opbasis}
\end{align}
where $q$ is a light quark field, which we take to be $u$ for our analysis.  If one were specifically interested in the modification of top quark couplings to SM gauge bosons, at one loop these operators generate an axial form factor for the top quark gauge coupling at $\mo(q^2/\Lambda^2)$ where $q$ is the boson momentum.

In the presence of other light quarks the limit on these $u \bar u \, t \bar t$ operators will get stronger, so the $u$-only case sets a conservative bound.  The $d$-only limit will be roughly half the scale of $u$-only, so a mix of light-quark production modes will strengthen the constraints on both flavors from their respective ``starting points.'' We do not consider any loop effects with these dimension-six operators, so the difference between the singlet and octet cases is purely due to a color factor, which we use to scale our bound from one channel to another.  These non-interfering operators contribute to the overall cross-section at $\mo(1/\Lambda^4)$.  At this order we become sensitive to interfering, dimension-eight operators.  As Ref.~\cite{Delaunay:2011gv} notes, by naive dimensional analysis, dimension-eight operators receive additional $1/16\pi^2$ suppression.  A particular source of QCD-interfering, dimension-eight operators is chirality-flipping, meaning that we have $m_t/\Lambda$ suppression.
  Thus, there are many new physics scenarios where one can safely neglect even higher dimension operators.  

\subsection{Analysis}
\label{subsec:ffalys}

To generate our new physics events, we added the operators in Eq.~\ref{eq:opbasis} to the SM in FeynRules \cite{Christensen:2008py} and generated 100K (SM + dimension-six operator) $t \bar t$ events in the $\ell$+jets channel in MadGraph 5 v$1.5.9$ \cite{Alwall:2011uj} run with Pythia 6.4 \cite{Sjostrand:2006za} for the 14 TeV LHC.  Background samples were run with Delphes 3.09~\cite{delphes} and we considered $t \bar t$, $W$+jets, $Z$+jets, $t$-channel single top quark, and $tW$.  An additional source of background comes from dijet events with various fakes.  We did not attempt to study this in detail, but we note that in the $\ell$+jets study of the $t \bar t$ charge asymmetry at ATLAS \cite{ATLAS:2012an}, contributions from dijet events were well beneath the other considered backgrounds.  

Due to the absence of interference with QCD, the operators we consider
only add $t \bar t$ events, and will do so preferentially for harder
momenta.  To set our limit, we measure the ratio of $\ttbar$ events
between a ``high'' and ``low'' bin, where we consider either $h_T$
(scalar sum of $p_T$ of all visible top quark decay products and \met) or
\mtt\ for a top quark reconstruction algorithm we detail below.  We can
parametrize our high-to-low ratio as
\begin{align}
r &= \frac{a + C^2 \,b }{d + C^2 \, f} \nonumber \\
  &= \frac{r_0 + C^2 \, B}{1 + C^2 \, F},
\label{eq:ratparam}
\end{align}
where $a,\,d$ are the SM contributions to the high and low bins, $C^2 b, C^2 f$ are those from new physics, $C$ is the dimensionful prefactor of our higher-dimension operator, and we obtain the second line from the first by dividing every term by $d$.  Thus, $r_0$ is expected ratio for SM $t \bar t$ $\ell$+jets events.  To set a limit, we need to understand how this observable scales with $C$.  This can be done by determining $r$ for three separate choices of $C$.  For simplicity, we take $C=0$ as one of these, giving us a determination of $r_0$.  The other points are chosen near the limit of perturbativity in order to give a large sample of new physics events to control statistical fluctuations.  They are $C = 3.15/(1 {\rm TeV})^2$ and $C = 6.3/(1 {\rm TeV})^2$.

We begin by turning on a single operator, $\mo_{AV}^{8}$.  By the end of Monte Carlo processing, events are showered, hadronized, had detector effects and pileup applied through Delphes,\footnote{We take a pileup of $\mu=50$ and scale from a luminosity of 300~fb$^{-1}$ to 3 ab$^{-1}$ purely through statistics.} and clustered into jets.  We apply various $p_T$ and $\eta$ cuts for event selection,\footnote{We demand that jets have $p_T >$ 35 GeV (leptons $>$ 25 GeV) and $\eta \, <$ 2.7 for jets and leptons.  Additionally, we require \met $>$ 30 GeV.} demand that there is only a single charged lepton, at least four jets that meet our criteria, and at least one $b$-tag.  Our analyses fall into two basic types.  In the first, we attempt no top quark reconstruction and only use transverse, measured quantities.  For this case, events are placed in the high/low bin depending on their $h_T$.  

In the second scenario, we cluster objects into top quarks and $W$s, taking the permutation of jets and neutrino $p_z$ solution that gives the best $\chi^2$ for $m_t$ and $m_W$.  As an additional criteria, we discard events that give complex $p_z$.  For our candidate jets, we first take any with $b$-tags and then the hardest untagged jets until we get to four total.  When clustering jets into heavy objects, if there are two or more $b$-tags, we demand at least one $b$ per top quark, and if there is one $b$ assigned to the harmonically decaying top quark, we form the $W$ out of the other, untagged jets. Upon reconstructing the top quarks, we determine $r$ for both $h_T$ and \mtt.  We also consider cuts on the invariant masses of the hadronic $W$ ($m_W^{had} \in [50,\,110]$) and $t$ ($m_t^{had} \in [125,225]$) to reduce backgrounds and improve event quality.  

To set a limit on our dimension-six operator, we perform a $CL_s$ analysis \cite{Read:2002hq}, following the procedure of \cite{Kilic:2012kw}.  We convolve Gaussian systematic errors with Poisson statistical fluctuations, using the ratio in Eq.~\ref{eq:ratparam} as our test statistic.  For systematic uncertainty, we use a modified version of the values in \cite{Aad:2013nca}), which reports both a ``resolved'' and ``boosted'' scenario, which take for our low and high bins, respectively.  The NNLO calculation reported in \cite{Czakon:2013xaa} allows us to reduce the error quoted by ATLAS.  We thus assume a 10\% uncertainty on background in the low bin, 20\% in the high bin, and 10\% uncertainty on the signal, taking all errors to be uncorrelated.\footnote{Replacing the theoretical uncertainties of \cite{Aad:2013nca} with those in \cite{Czakon:2013xaa} would give us respective uncertainties of 13.4\% and 23.4\% for the low and high bins.  We thus assume some improvement in these numbers by the time the LHC has 
many 100~fb$^{-1}$ for analysis.  Our 
uncertainty on the signal comes from those quoted for the $Z'$ analysis in \cite{Aad:2013nca}, which were 11.2\% and 7.1\%, which have modified slightly for simplification.}  We present limits for $(C_{AV}^8)^{-1/2}$ for our best choices for bin variable, bin boundary, and cuts in Table \ref{tbl:lims}.  We placed the boundary between the high and low bins for $h_T$ and \mtt\ at 750 GeV, 1 TeV, and 1.25 TeV.  Even higher values begin to lose statistics at 300~fb$^{-1}$ and we may become sensitive to additional systematic effects.  
\begin{table}[htdp]
\begin{center}
\begin{tabular}{|c|c|c|c|}
\hline
bin variable & low/high border &reconstruction cuts & 300~fb$^{-1}$ \\ \hline
$h_T$ & 1 TeV & $m_t$ & 910 GeV \\ \hline
$h_T$ & 1 TeV & $m_t$,\,$m_W$ & 910 GeV  \\ \hline
$h_T$ & 1.25 TeV & - & 950 GeV  \\ \hline
\mtt & 1.25 TeV & $m_t$ & 670 GeV   \\ \hline
\end{tabular}
\end{center}
\caption{Limits (2$\sigma$ with $CL_s$) on the effective scale of new physics, $(C_{AV}^8)^{-1/2}$, for 300~fb$^{-1}$ of LHC14 data.  They are determined by using the ratio $r$ ({\it cf.~}Eq.~\ref{eq:ratparam}) for either $h_T$ or \mtt\ bins as a test statistic.  Besides our basic analysis cuts, in some cases we require $m_t^{had} \in [125,225]$ or $m_W^{had} \in [50,\,110]$.}
\label{tbl:lims}
\end{table}
%

\subsection{Four-quark operator conclusion}
\label{subsec:fqoConc}

We have found that with very simple analysis techniques we can push non-interfering dimension-six operators to the TeV-scale in the near future with the LHC.  In fact, our strongest bound came in the case of no top quark reconstruction, using only the $p_{\rm T}$ values of decay products that correspond to a $t \bar t$ pair decaying in the $l$+jets channel. Our bounds are systematics limited, and thus an increased luminosity would have a marginal effect.  A higher energy machine would of course push our limits farther provided one could meet the challenges of working in a regime of ultra-boosted top quarks.  

In Eq.~\ref{eq:opbasis} we listed eight operators, though we have only studied one in detail.  As mentioned above, we can convert limits from octet to singlet operators by using the appropriate color factor.  The corresponding limits for $(C_{AV}^1)^{-1/2}$ can thus be obtained from Table \ref{tbl:lims} by multiplying the results by $(3/\sqrt{2})^{-1/2}$.  As for the others, adding them to the SM yields very similar changes to the overall cross section.  Thus, we expect the limits to be comparable.  If we were to observe a significant modification to the overall $t \bar t$ rate, it is an interesting question how we would disentangle the operators involved.

\section{Measurement of the $Wtb$ coupling in single top quark production}
\label{vtb} 
The measurement of the $Wtb$ coupling, or more generally $c^W_L$ in Eq.~\ref{eqn:anom} (assuming there are no CP-violating interactions and the coupling is of type V-A, i.e. $c^W_R=d^W_L=d^W_R=0$), can be performed with no assumptions on the number of quark families or the unitarity of the quark-mixing matrix by counting the rate of production of single top quarks. 
The current values of the coupling strength $c^W_L$ extracted from single top quark production are $c^W_L=1.04^{+0.10}_{-0.11}$ from ATLAS (5.8~fb$^{-1}$)~\cite{atlas-tchan} and $c^W_L=0.96 \pm 0.08$(syst.)$\pm 0.02$(theory) from CMS (5.0~fb$^{-1}$)~\cite{cms-tchan}. The relative uncertainty of these measurements is 11\% and 7\% respectively. 
The Tevatron combined measurement with 2.3 and 3.2~fb$^{-1}$ obtains a similar sensitivity and measures $c^W_L=0.88 \pm 0.07$, with a relative uncertainty of 8\%~\cite{vtb-tev}.  
Additional information can be obtained from the study of the decays of top quarks produced in pairs, which give a more precise determination of $c^W_L=1.011^{+0.018}_{-0.017}$ with a 1.8\% relative uncertainty for 17~fb$^{-1}$~\cite{cms-vtb-rratio}. However, these results from top quark pair production and decay assume that there are only three generations and the CKM matrix is unitary.

A study has been performed on the measurement of $t$-channel single top quark production cross section at the LHC with 14~TeV center of mass energy, with pileup conditions of $\mu=50$ and $\mu=140$~\cite{TchanWhitePaper}. Signal samples were generated with MadGraph~\cite{Alwall:2011uj} and Pythia~\cite{Sjostrand:2006za} for hadronization, and passed through a generic detector simulation in Delphes v3.09~\cite{delphes}. The generated events at 14~TeV correspond to 300~fb$^{-1}$ with pileup of 50 interactions, and 3000~fb$^{-1}$ with pileup of 140 interactions on average. Background samples corresponding to $t\bar{t}$, $W/Z$+jets, and multijet were obtained from the Snowmass simulation~\cite{MCtwiki}. A cut-based analysis was performed requiring at least two jets with $p_T>50$~GeV, one of which is $b$-tagged, and one forward jet with $|\eta| > 3$.  Additional selection criteria to enhance the signal require the reconstructed top quark mass to be between 160 and 180~GeV and $\cos(lj,\ell)>0$ which measures the 
top quark polarization from the angle between the light jet and the charged lepton. The analysis expects around 1,000 signal events in 300~fb$^{-1}$, with a signal to background ratio of 2:1. Assuming a 10\% uncertainty on the background yield, the estimate from this study on the relative uncertainty in $c^W_L$ is 2.5\%. A similar $c^W_L$ uncertainty is also obtained for 3000~fb$^{-1}$ and $\mu=140$. 

All these measurements of $c^W_L$ from the single top quark cross section allow the strength of the $Wtb$ coupling to deviate from one in the production of the top quark, but force the SM value for the decay. This is a reasonable framework given that $V_{tb} \gg V_{td}, V_{ts}$, so if a new heavy quark exists the $V_{tb}$ value could be smaller than 1 but the decay would still proceed through $t \to Wb$ almost 100\% of the time. 
Limits on a possible fourth generation of heavy $b^\prime$ quarks has been set from direct searches in the single top quark final state ($m_{b^\prime} > 870$~GeV at 95\% CL)~\cite{atlas-bprime}, or indirectly from the measurement of the width of the top quark ($V_{tb^\prime} < 0.59$ at 95\% CL)~\cite{d0-width}. 
D0 has performed searches for anomalous couplings in the $Wtb$ vertex $c^W_R, d^W_L, d^W_R$~\cite{d0-anom} by combining the sensitivity of the $t$-channel production and the $W$-helicity measurement from the $t\bar{t}$ decays. The results are consistent with $c^W_R=d^W_L=d^W_R=0$ and $c^W_L=1$, but the large statistics of the LHC will allow to constrain these couplings (and the corresponding operators described in Sec.~\ref{sec:fourfermi}) even more. Indeed, the sensitivity of the LHC experiments to anomalous $Wtb$ couplings in the single top production has been studied in~\cite{Bach:2012fb}. This analysis shows the kinematic effects on selection acceptances in $s$ and $t$ channel final states coming from the presence of different anomalous couplings. The sensitivity to different combinations of $c^W_R, d^W_L, d^W_R$ is studied for 10~fb$^{-1}$ at 14~TeV, taking also into account the interference with a charged-current four-fermion interaction. The results show that it is possible to measure simultaneously within 1$\sigma$ values of $|d^W_L|< 0.2$ and $|d^W_R|<0.15$, assuming $c^W_L=1$, and $c^W_R=0$. 

\section{Coupling of the Top quark to charge zero vector bosons}
\label{sec:ttzero}

Two decades after the discovery of the top quark most of its couplings
are still to be measured. The $\ttbar \gamma$ ($\ttbar Z$) coupling
may be determined via analysis of direct production of $\ttbar$ in
association with a $\gamma$ ($Z$). So far only evidence for these
processes has been
reported~\cite{cdf_ttgamma,Chatrchyan:2013qca}. Unlike at a linear
$e^+e^-$ collider~\cite{Asner:2013hla,Amjad:2013tlv}, the LHC's
capability of associated $\ttbar\gamma$ and $\ttbar Z$ production has
the advantage that the $\ttbar\gamma$ and $\ttbar Z$ couplings are not
entangled.

\subsection{$\ttbar \gamma$}
\label{s:ttg}

The coupling of the top quark to photons (and therefore the
$\ttbar\gamma$ production cross section) is sensitive to the electric
charge of the top quark. In addition, the $\ttg$ production is an
important control sample for the $\ttbar H,~H\rightarrow\gamma\gamma$
process. Moreover, new physics connected with EWSB can manifest itself
in top precision observables. For instance, the $\ttbar
\gamma(\gamma)$ yield can put constraints on excited top quark
($t^*\rightarrow t\gamma$) production, especially in case of
predominantly electromagnetic decays of the $t^*$, when $t^*$
de-excites via radiating a photon (rather than a gluon as discussed in
Ref.~\cite{Stirling:2011ya}).

The distinction between the radiative $\ttbar$ production
($pp\rightarrow\ttbar\gamma$) and radiative $\ttbar$ decay is critical
for determination of $\ttg$ couplings. To select events with photon
emission from top quarks, the photon radiation from the $W$ and its
decay products, as well as from the $b$ quarks and from the initial-
and final-state partons should be suppressed, as detailed in
Ref.~\cite{tt_couplings_baur}. In addition, the event should be
consistent with the radiative $\ttbar$ production, which can be
checked by performing top quark and top quark + photon mass
reconstruction. However, the analyses performed so far do not have the
sensitivity to disentangle the radiative $\ttbar$ production decay and
 to measure the $\ttg$ couplings.

The first evidence for $\ttg$ production was presented by the CDF
collaboration~\cite{cdf_ttgamma}. The ratio
$R=\sigma_{\ttg}/\sigma_{\ttbar}$ was measured to be $0.024 \pm 0.009$
in agreement with the SM expectation of $0.024 \pm 0.005$ for a charge
2/3 top quark and a photon $p_T > 6$ GeV (using a photon $p_T > 10$
GeV in the reconstruction). The uncertainty of the measurement is
limited by statistics.

The ATLAS collaboration performed the first $\ttbar\gamma$ production
cross section measurement~\cite{atlas_ttgamma} at the LHC in 1.04
fb$^{-1}$ of data at $\sqrt{s}=7$ TeV, $\sigma_{\ttbar\gamma}=2.0\pm
0.5~(stat.)\pm 0.7~(syst.)\pm 0.08~(lumi.)$~pb for a photon $p_T$
threshold of 8 GeV (using a photon $p_T > 15$ GeV in the
reconstruction). The significance of the measurement is 2.7~$\sigma$,
and the measured $\sigma_{\ttg}$ is in agreement with the SM
expectation of $2.1\pm 0.4$~pb for a charge 2/3 top quark. The
uncertainty of the measurement is dominated by the photon
identification efficiency, the initial and final state radiation
modeling and the jet energy scale systematics. Larger 7 and 8 TeV
datasets will allow to significantly reduce both systematic and
statistical uncertainties.

The LO $\ttbar\gamma$ production cross section increases by a factor
of 5 from 7 TeV to 14 TeV center-of-mass energy for photons with
$p_T>20$~GeV~\cite{Stelzer:1994ta,Maltoni:2002qb}. With the much
larger $\ttg$ statistics expected after the 2013-2014 shutdown,
analysis of the $\ttg$ couplings is feasible. The strategy is to
isolate events with photon emission from top quarks, as well as to
perform a ratio $R=\sigma_{\ttg}/\sigma_{\ttbar}$ measurement to
reduce $\ttbar$-related systematic uncertainties. The limiting factors
are the photon identification in busy $\ttg$ environment in high
pile-up conditions, as well as proper selection of events with
radiative top-quark production.

At the LHC, with 300 $\invfb$ several thousand signal events are
expected, therefore precise determination of the $\ttbar\gamma$
couplings is possible using the lepton plus jets
channel~\cite{tt_couplings_baur}. In addition, the measurement should
be performed using the dilepton channel that is expected to provide a
smaller systematic uncertainty (due to fewer jets in the event) but a
slightly larger statistical uncertainty compared to the lepton plus
jets channel. With 3000 $\invfb$ of data expected at the High
Luminosity (HL) LHC, differential measurements of $\ttbar\gamma$
couplings (for instance, as a function of photon $p_T$) as well as
differential $\ttbar\gamma$ cross section measurements, can be
performed.

\subsection{$\ttbar Z$}
\label{s:ttz}

The associated $\ttbar Z$ production is directly sensitive to $\ttbar
Z$ couplings. New physics~\cite{Randall:1999ee,Agashe:2004rs} can
modify the $\ttbar Z$ vertex. These effects can be studied by
measuring differential $\sigma_{\ttbar Z}$ as a function of $Z$
transverse momentum and as a function of $\Delta\phi(\ell^+,\ell^-)$
for the leptons coming from $Z\rightarrow\ell^+\ell^-$. However, the
inclusive $\sigma_{\ttbar Z}$ should be measured first.

The associated production of $\ttbar$ and $Z$ or $W$ bosons has been
measured by CMS in 5 fb$^{-1}$ of pp collisions at
$\sqrt{s}=7~$TeV~\cite{Chatrchyan:2013qca}. The measurements exploit
the fact that SM events with two prompt same-sign isolated leptons in
the final state, as well as trilepton events, are very rare. The
background contribution due to jets misidentified as leptons is
estimated using a data-driven procedure.

A direct measurement of the $\ttbar Z$ cross section $\sigma_{\ttbar
  Z} = 0.28^{+0.14}_{-0.11} (stat.)^{+0.06}_{-0.03} (syst.)$~pb is
obtained in the trilepton channel, where 9 events are observed with
the expected background of 3.2 $\pm$ 0.8 events. The signal
significance is 3.3~$\sigma$ from the background hypothesis. In the
dilepton channel a total of 16 events is selected in the data,
compared to a background expectation of $9.2\pm 2.6$ events. The
presence of a $\ttbar V$ ($V=W, Z$) signal is established with a
significance of 3.0~$\sigma$. The $\ttbar V$ process cross section is
measured to be $\sigma_{\ttbar V} = 0.43^{+0.17}_{-0.15} (stat.)
^{+0.09}_{-0.07} (syst.)$~pb. Both $\sigma_{\ttbar V}$ and
$\sigma_{\ttbar Z}$ measurements are in agreement with the NLO
predictions~\cite{ttz_baur}. The ATLAS collaboration also performed
$\ttbar Z$ analysis~\cite{ttz_atlas}, using a much tighter selection
to suppress backgrounds, that resulted in observing one event in data
and setting a limit $\sigma_{\ttbar Z} < 0.71$~pb at 95\% CL
consistent with the CMS measurement.

The uncertainties of the $\ttbar Z$ measurements up to date are
dominated by statistics. The LO $\ttbar Z$ production cross section
increases by a factor of $\sim$ 1.4 from 7 TeV to 8 TeV center-of-mass
energy~\cite{Stelzer:1994ta,Maltoni:2002qb}, so the expected decrease
in the statistical uncertainty with the dataset collected in 2012 is a
factor of 2.5. Therefore, statistical uncertainties are expected to
dominate the $\sigma_{\ttbar V}$ and $\sigma_{\ttbar Z}$ measurements
performed with 2012 data.

The LO $\ttbar Z$ production cross section increases by roughly an
order of magnitude from 7 TeV to 14 TeV center-of-mass
energy~\cite{Stelzer:1994ta,Maltoni:2002qb}, therefore precise
measurements of the $\ttbar Z$ production cross section can be
performed after the 2013-2014 shutdown.  According to
Ref.~\cite{tt_couplings_baur}, with 300 $\invfb$ of 14 TeV collisions
data, the $\ttbar Z$ vector (axial vector) coupling can be determined
with an uncertainty of 45 -- 85\% (15 -- 20\%), whereas the
dimension-five dipole form factors can be measured with a precision of
approximately 50\%. For 3000 $\invfb$ the limits are expected to
further improve.


\section{Top Yukawa Coupling}
\label{sec:yukawa}
Of the current $t\bar{t}H$ searches, CMS sets the most stringent
limits, with an observed (expected) combined upper limit at 95\%
confidence level (C.L.) of 3.4$\times$ (2.7$\times$) the SM
expectation~\cite{cmstth} using the full 2012 and 2011 LHC data sets.
Multiple analyses enter into this combination - analyses optimized for
$H \rightarrow b\bar{b}$ in the $t\bar{t}$ lepton+jets and dilepton
channels, a single lepton $H \tau \tau$ analysis and an inclusive
single lepton $H \rightarrow \gamma \gamma$ analysis.  Unlike in the
hadronic Higgs decays (or Higgs decays with multiple neutrinos), the
narrow reconstructed width of the diphoton resonances allows for a
sideband estimate of the expected background~\cite{cmstthphoton}.
Both hadronic and inclusive leptonic channels are evaluated. The
hadronic channel requires five or more jets in the event, at least
one of which must be $b$-tagged, and no charged leptons. The leptonic
channel requires at least two jets, at least one of which must be
$b$-tagged, and at least one electron or muon.  The expected
(observed) limits are 5.3$\times$(5.4$\times$) the SM expectation. 
The observed signal strength is +0.21$^{+2.18}_{-1.46}$.

ATLAS also recently finished a diphoton analysis in the hadronic and
leptonic channels~\cite{atlastthphoton}.  In the hadronic channel,
five or more jets (at least two of which must be $b$-tagged) are
required.  In the leptonic channel, a single $b$-tagged jet and at
least one lepton are required, along with a small MET requirement.
The combined upper limits on $t\bar{t}H$ production are 5.3$\times$
the SM prediction, with an expected limit of 6.4.  In both the ATLAS
and CMS diphoton analyses, systematic uncertainties worsen expected
limits by only $\sim 6\%$ due to the low statistics available both for
the signal as well as for the background estimation from the
sidebands. The largest systematic uncertainties are due to the unknown
heavy flavor and large $\njet$ contributions from non-$t\bar{t}H$
Higgs boson production, in particular glue-glue fusion. These
uncertainties are in particular large for the all-hadronic analyses,
where large numbers of jets and several $b$-tags are required.

Prospects for the $t\bar{t}H$ channel with Higgs decay to diphotons and
3000~fb$^{-1}$ of data were also studied in the context of Snowmass
with the high-pileup $<\mu>=140$ samples. Events are required to
contain two photons with $p_T$ above 30~GeV, at least one lepton with
$p_T$ above 25~GeV, at least two $b$-tags (and at least three jets in
total), and at least 30~GeV of MET. In order to smooth out statistical
fluctuations in the simulation, a jet-photon fake rate is estimated
from photon+jet events, and applied randomly to jets in the
simulation. In the absence of truth information, the fake rate is
derived by looking at the ratio of single photon vs diphoton events in
the sample and dividing by the number of jets in the single photon
events, and is estimated to be 0.0046.  The direction of the fake
photon is taken to be the direction of the jet, and the $p_T$ is
randomly assigned using the the highest-pileup jet-to-fake photon
$p_T$ parameterization from the ATLAS European Strategy performance
studies~\cite{atlasesperformance}. Simulated events are looped over
5000 times with the above procedure to further smooth out statistical
fluctuations.  The backgrounds that dominate are $W$ and $Z$ boson
production in association with extra jets and/or photons, and
$t\bar{t}$ itself in conjunction with extra jets and/or photons. Due
to technical limitations, the $t\bar{t}$ +0 photon and $W/Z$ boson
+0/1 photon samples are generated at 13 TeV and not at 14 TeV
(however, the samples are normalized to 14 TeV cross sections). The
$t\bar{t}$ +1/2 photon and $W/Z$ boson +2 photon samples are the
correct center of mass energy.  The background sideband is fit to an
exponential.

\begin{figure}[!htp]
  \centering
  \includegraphics[width=0.47\textwidth]{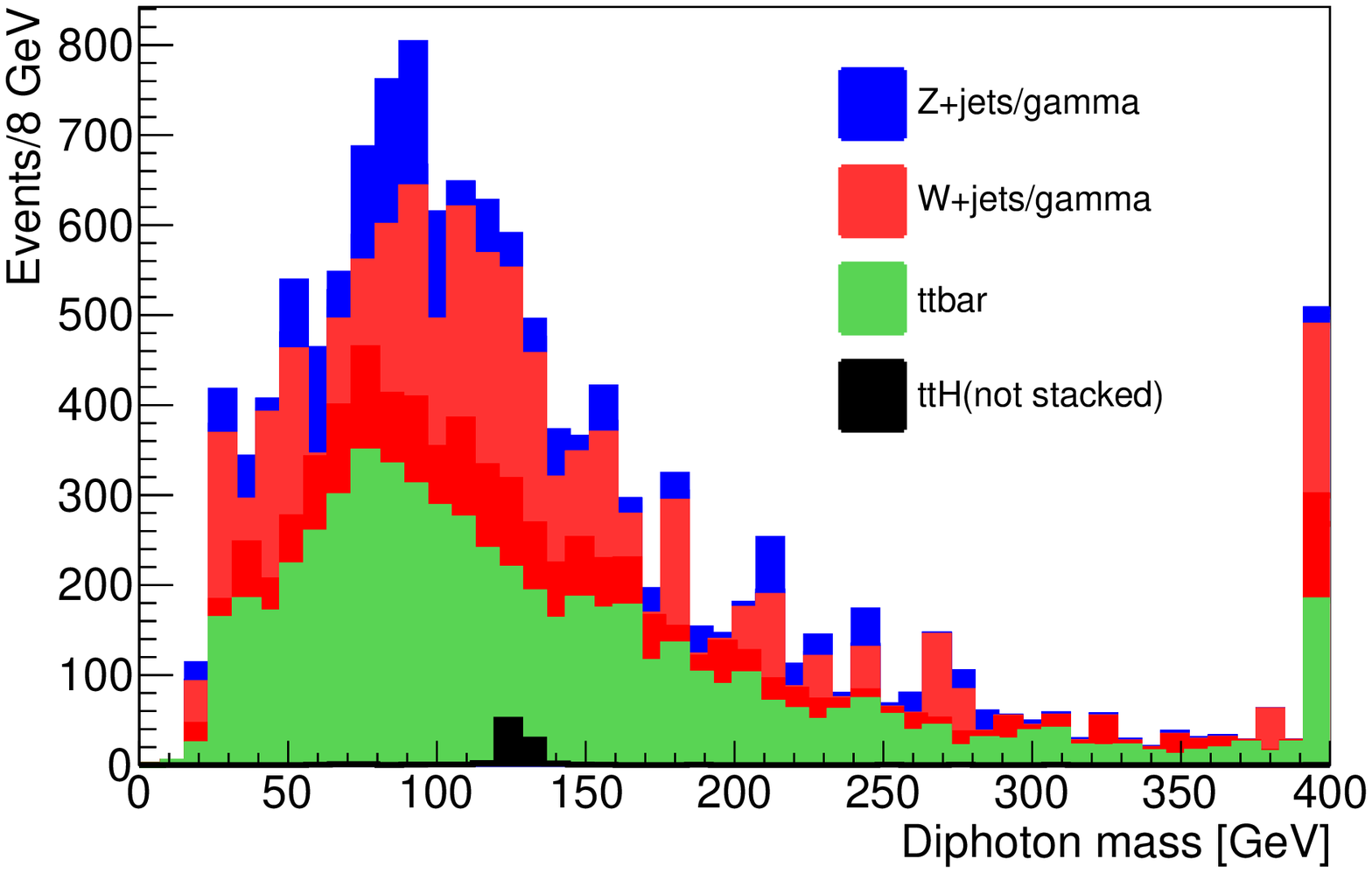}
  \includegraphics[width=0.47\textwidth]{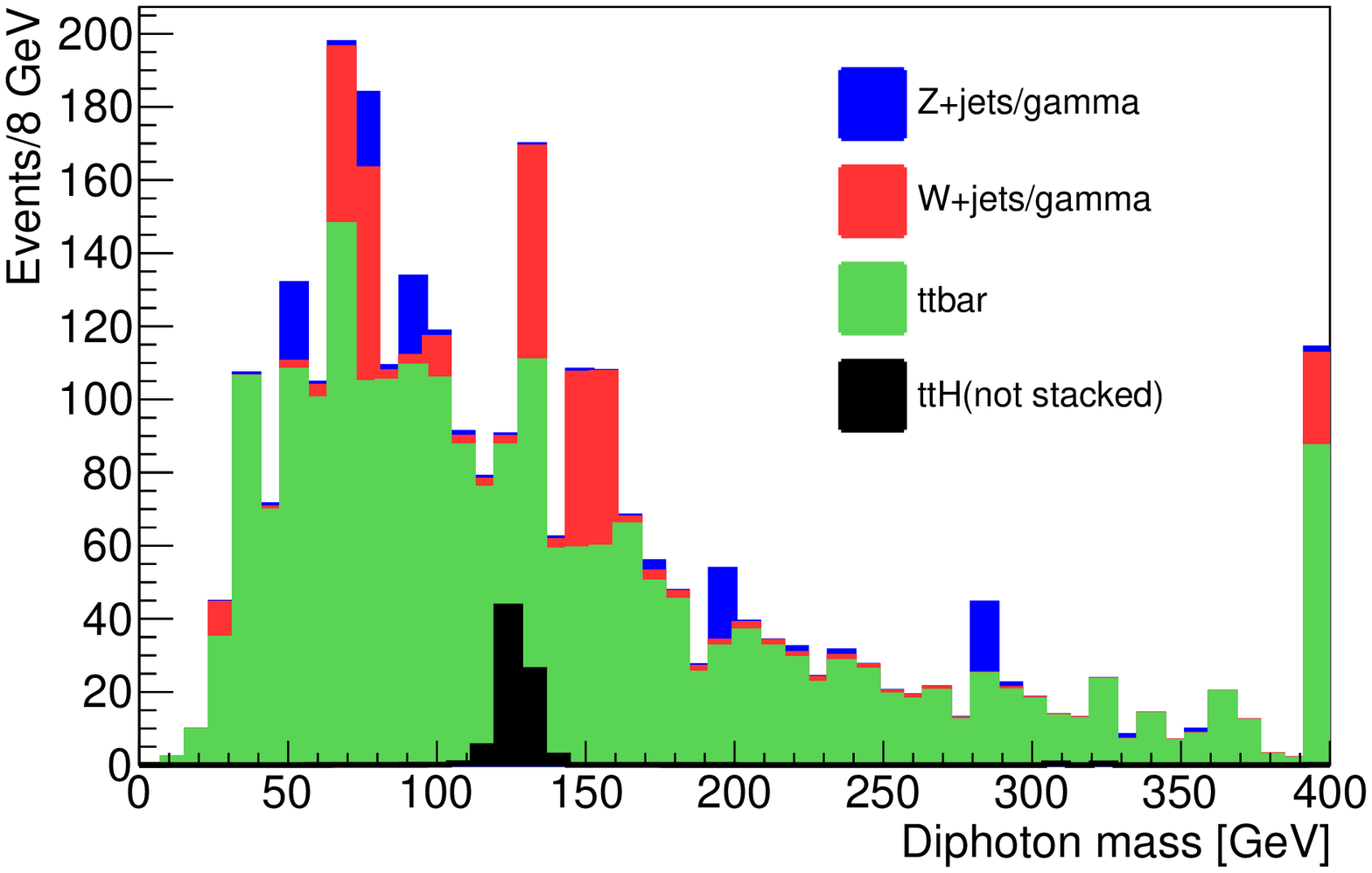}
  \caption{Diphoton mass spectrum expected at 14~TeV after all selection cuts and the 1 exclusive $b$-tag selection (left) and 2 $b$-tag selection (right) for 3 ab$^{-1}$ of data. The plots are stack plots except for the signal, which is plotted at the bottom of the canvas on top of the backgrounds.}
  \label{fig:diphotonttH}
\end{figure}

With 3 ab$^{-1}$ of data and requiring two or more
$b$-tags, $S = 58$ and $B = 122$, giving $S/\sqrt{B} = 5.2$, not very
different from the values obtained by ATLAS in European Strategy
studies~\cite{atlases}. The requirement of two $b$ tags significantly
reduces the contributions from the electroweak backgrounds.  Similar
limits but with a slightly worse $S/\sqrt{B}$~(S=67, B=639) are
obtained by using an exclusive single-tag selection, in which the
signal increases, but the backgrounds increase substantially.
Combining the two channels gives $S/\sqrt{B}$ = 5.9.  Pushing the
$\njet$ requirement to 4 or more jets was also evaluated, but was
shown to give worse performance. Other cuts studied included
variations on the MET, the lepton $p_T$ and the jet $p_T$.
Figure~\ref{fig:diphotonttH} shows the diphoton mass spectrum after
the optimized selection. It is clear that in this channel,
300~fb$^{-1}$ of data is not enough to see a signal, but that it is an
important one, as the excellent mass resolution allows for an
unambiguous observation of the Higgs boson resonance. In addition, the
diphoton channel remains an important candidate for searches for new
physics in the Higgs sector at the LHC, as large data sets and the
unambiguous assignment of photons to the Higgs boson may allow for
tests of the top quark-Higgs boson vertex, including whether there is
any CP-odd component to the Higgs boson coupling to top quarks~\cite{cpvertex}. Such
measurements can also be made in the $H\rightarrow\mu\mu$ channel, but
even with 3000~fb$^{-1}$ they are likely to be statistically
limited~\cite{tthmumu}.

The $b\bar{b}$ analyses from CMS are divided into several bins
depending on whether the event contains one or two charged leptons,
the number of jets and the number of $b$-tags. Each of these
sub-analyses have separately trained boosted decision trees (BDTs)
that use 4-10 variables to separate signal and background.  Variables
include event shape information, jet kinematics, MET, $b$-tagging
information (to separate out $t\bar{t}$ + light flavor jets from
$t\bar{t}$ + heavy flavor jets), angular information and kinematics
associated with attempts to reconstruct the Higgs boson mass from the
$t\bar{t}H$ hypothesis. The dominant uncertainties are 50\% rate
uncertainties applied independently to $t\bar{t}+b\bar{b}$,
$t\bar{t}+c\bar{c}$ and $t\bar{t}+b$. The observed (median expected)
limit is 5.2 (4.1)$\times$ the SM expectation. The best-fit value for
the signal strength is +0.85$^{+2.47}_{-2.41}$.  The ATLAS $b\bar{b}$
measurement~\cite{atlasbb} uses the full 7 TeV dataset and focuses on
the single lepton channel. Events are divided into different
categories based on the number of jets and $b$-tags. A fit to the
$t\bar{t}H$ hypothesis is made on events with six or more jets, with
the reconstructed Higgs boson mass the observable of interest. Events
with fewer than six jets use the scalar sum of the hadronic jet $p_T$
in the event as the discriminating variable. The observed (expected)
limit is 13.1 (10.5)$\times$ the SM value. As in the CMS measurement,
the dominant systematic uncertainty is on the modeling of extra heavy
flavor produced in association with $t\bar{t}$ events.


CMS and ATLAS have made first studies of production of top-quark pairs
together with heavy-flavor (HF) quarks, which is the main irreducible
background to $\ttbar + H$, $H \rightarrow b\bar{b}$. CMS performed a
measurement~\cite{cmsratio} of the ratio of production cross sections:
$\sigma(t\bar{t}b\bar{b})/\sigma(t\bar{t} jj)$. ATLAS studied the
production of $\ttbar + b + X$ or $\ttbar + c + X$, collectively
referred to as $\ttbar$ + HF by measuring the ratio of production
cross sections $\sigma(t\bar{t}+{\rm HF})/\sigma(t\bar{t}j)$~\cite{atlasratio}. 
Both measurements are made in a given visible phase space. The results
are larger than predictions, though with big uncertainties. Increased
luminosities would potentially improve these
measurements, which would significantly improve the sensitivity of
these channels. In addition, the technique used by ATLAS will allow
separate measurements of $\ttbar + b + X$ and $\ttbar + c + X$ to
reduce systematic uncertainty on the modeling of extra heavy flavor
produced in association with $t\bar{t}$ events.  Long-term prospects
for the $H \rightarrow b\bar{b}$ channel are being pursued in the
Snowmass high-energy frontier Higgs group.

Finally, the best channel for $t\bar{t}H$ studies at hadronic colliders
might be the ditau channel~\cite{PhysRevD.86.073009}. The $t\bar{t}$
system is required to decay to the single lepton channel, and at least
one of the taus from the Higgs decay is required to decay
hadronically. Events are then split into whether both taus decay
hadronically or one decays leptonically. The charged lepton is
required to have $p_T > $~25 GeV, and the two taus are required to be
oppositely charged with $p_T > $~15GeV.  In addition, at least 4 jets
are required, and events consistent with coming from a $Z$ boson are
removed. With only 100~fb$^{-1}$, $S/\sqrt{B} = 3.7(2.1)$ in the
$\tau_h\tau_l(\tau_h\tau_h)$ channel. These numbers improve to
$S/\sqrt{B} = 6.4(3.7)$ in the $\tau_h\tau_l(\tau_h\tau_h)$ channel
with 300~fb$^{-1}$ of data, indicating the potential for better than
20\% cross section accuracy with only 300~fb$^{-1}$ of data. The
combined CMS result also examines the ditau channel by using BDTs,
setting observed (expected) limits of 13.2 (14.2)$\times$ the SM
expectation.

Other areas of active study for $t\bar{t}H$ searches also include the
multiple channels in which $H\rightarrow WW$, which can lead to
various signatures, including same-sign dileptons, trileptons, four
leptons, and multiple leptons with additional hadronic taus. These are
also being studied by the Higgs group in the context of Snowmass.

\section{Conclusions}
\label{sec:conc}
Studies of the the top quark and its couplings at the LHC will continue to yield insight into the Standard Model and potential new physics for quite some time. Understanding and testing the top quark Yukawa coupling and searching for new physics with other top quark couplings are an important part of the future LHC program.


\bibliography{lit}
\bibliographystyle{jhep}

\end{document}